\documentclass{elsart5p}
\usepackage{psfrag}
\usepackage{amssymb}
\usepackage{graphicx}
\begin{document}
\begin{frontmatter}

\title{Radio Detection of Particles from the Cosmos}

\vspace{4mm}
\author{
Andreas~Haungs}

\vspace{4mm}
\address{
Institut\ f\"ur Kernphysik, Forschungszentrum Karlsruhe, Germany\\
email: andreas.haungs@ik.fzk.de}
%\ead{andreas.haungs@ik.fzk.de}

\begin{abstract}
The ARENA08 conference, held in Rome in June 2008, gave an almost complete overview on 
applications of the new techniques radio and acoustic detection of electromagnetic showers 
generated by high-energy cosmic particles in air or in dielectric media. 
There are vast activities all over the world, and the progress is remarkable. This was 
displayed by more than 30 contributions to the conference related to the radio detection 
technique, only. This paper gives a short summary on the status of 'radio detection 
of particles from the cosmos' as presented at the conference. 
\end{abstract}

\end{frontmatter}

\section{Motivation}

\begin{figure}
\begin{center}
\includegraphics*[width=8.0cm]{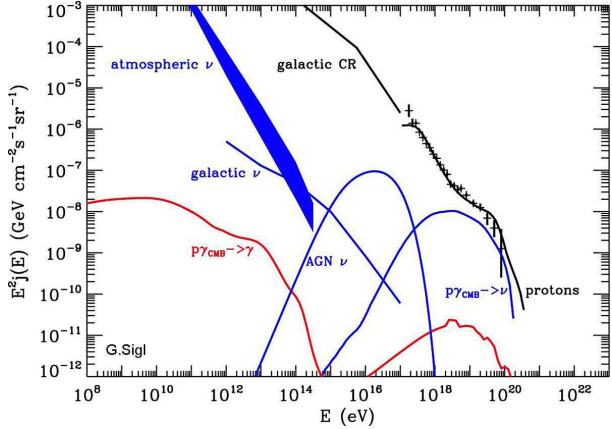}
\end{center}
\caption{Model fluxes of particles from the cosmos~\cite{Sigl06}. 
Shown are the primary cosmic ray fluxes (data and a model), 
the secondary gamma ray flux expected from proton interactions with 
the CMB and infrared background, and neutrino fluxes (atmospheric $\nu$, 
galactic $\nu$, AGN $\nu$, and GZK-$\nu$).}
\label{allflux}
\end{figure}

The new detection technique of radio observation of high-energy cosmic 
particles belongs to an innovative field of Astroparticle Physics: 
The study of high-energy cosmic rays, gamma rays, and neutrinos; i.e.~charged 
as well as neutral particles and nuclei from the cosmos. 
The goal is to get a coherent physical description of the high-energy, 
non-thermal Universe. 
The scientific questions address the understanding of cosmic accelerators and 
the origin of high-energy cosmic rays. 
The experimental challenge is to perform multi-messenger observations of 
astrophysical sources. 
These objects accelerate charged particles and other particles such as neutrinos 
or gamma rays are produced at reactions with the surrounding media.
In addition, due to the unavoidable interaction of very high-energy protons with 
photons of the microwave background (GZK-effect) secondary neutrinos and gamma
rays are produced (see fig.~\ref{allflux}).

The observation of particles at highest energies is a challenging task. 
As all cosmic messengers are rare and difficult to detect, the techniques 
employed require deployment of large area detectors and use of large volume 
natural detector media (atmosphere, sea water, ice, salt mines, etc.). 
Presently, the experimental field is entering a new era using large-scale 
detectors on the surface of the Earth, embedded in the icecap of the Antarctica, 
or using the deep-sea in the Mediterranean. 

With the recent results from the Pierre Auger Collaboration~\cite{Abraham07} of 
correlations between cosmic ray arrival directions and nearby Active Galactic Nuclei 
(AGN) a new window to the Universe has been opened.
AGN's could be the source of the highest energy cosmic rays, but definite 
conclusions can be only obtained when the statistical accuracy will be improved 
at the highest energies and when the information from different messengers 
(high-energy gamma rays, neutrinos, and hadrons) will be combined.

Classical techniques for the detection of cosmic rays are extensive-air-shower 
observations at ground level by operating large arrays of
detectors for charged particles, covering hundreds to thousands of square kilometres. 
Alternative optical techniques include observations of Cherenkov light in 
air, water, and ice, as well as observations of fluorescence light from 
nitrogen molecules in the atmosphere. 
These techniques are well advanced but become very expensive for the instrumentation of large areas. 
Therefore, if the observatories will be expanded the development of new efficient 
detection techniques is indispensable. 

Radio detection addresses a novel approach to unveil the origin, nature and 
propagation of cosmic rays, high-energy gammas, and neutrinos.
The approach can be suitable to develop new and cheaper detection techniques
with a large potential for application in the next generation of big experiments 
either in the Earth's atmosphere or in dielectric materials 
such as the ice of the polar caps, the salt of suitable mines and the lunar regolith. 
At sufficiently high energies the cosmic particles produce detectable signals in 
conveniently accessible radio frequency bands.
The general advantages of the radio technique lie in the fact that radio 
waves are comparatively easy to detect with cheap equipment. 
Radio waves propagate in a number of media without considerable attenuation. 
Therefore, radio signals can, in principle, be transported
over large distances in the atmosphere (several tens of kilometres) 
or through dielectric solids (several hundreds of meters). 
At various places in the world these techniques have been tested and are 
being used at relatively small scales. 
In the next few years the currently operated prototypes will be extended to first
large scale applications of the radio detection technique. 
 
\section{Radio detection of air showers}

\subsection{Charged cosmic rays}

\begin{figure}
\begin{center}
\includegraphics*[width=7.8cm]{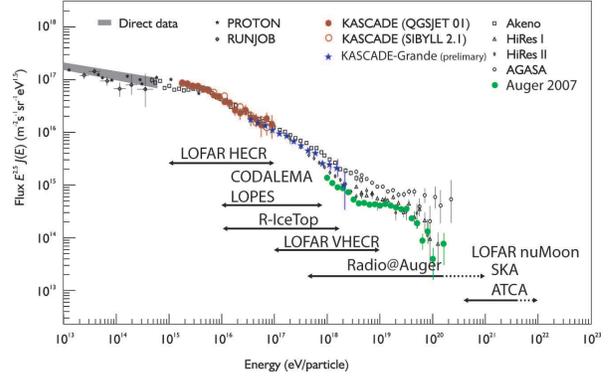}
\end{center}
\caption{The measured charged particle cosmic ray spectrum with indications 
for the sensitive energy range of various radio detection set-ups for air-shower 
observations.}
\label{crflux}
\end{figure}

In recent years, the radio detection of high-energy cosmic rays with the Earth's atmosphere as a target has been developed very rapidly and the proof-of-principle has been given by different experiments~\cite{Besson07}. 
The obtained result on the physics and the available infrastructure serve as milestones for the 
development of the next generation of experiments which will be able to investigate
in radio the whole high-energy spectrum of charged cosmic rays (see fig.~\ref{crflux}).

The detection principle of high-energy air showers is based on the idea that
electron-positron pairs gyrate in the Earth's magnetic field and produce radio 
pulses by synchrotron emission.
During the shower development the electrons  
are concentrated in a thin shower disk ($<2\,$m), 
which is smaller than one wavelength (at $100\,$MHz) of the 
emitted radio wave.
This situation provides the coherent emission of the radio signal.
There are different descriptions of this geomagnetic effect 
on charged particles moving in the atmosphere which led to various approaches 
of detailed Monte-Carlo simulations of the emission process (see~\cite{huege08} and 
references therein). 
In a next step, with increasing importance, a detailed comparison of the 
calculated signals using end-to-end simulations, starting from the showers themselves, 
including the details of the sensors and the sensor electronics 
is envisaged~\cite{Fliescher08}.

The advantage to perform coincidence measurements with existing well-calibrated
air-shower arrays of particle detectors made an unambiguously proof-of-principle 
of the radio detection of air-showers possible~\cite{Falcke05}.  
Meanwhile, various experiments have positive detection of air showers in the radio 
frequency range of $40-80\,$MHz.
Several thousand events have been analyzed in terms of features of the 
radio signal emitted in air shower. 

\begin{figure}
\begin{center}
\includegraphics*[width=7cm]{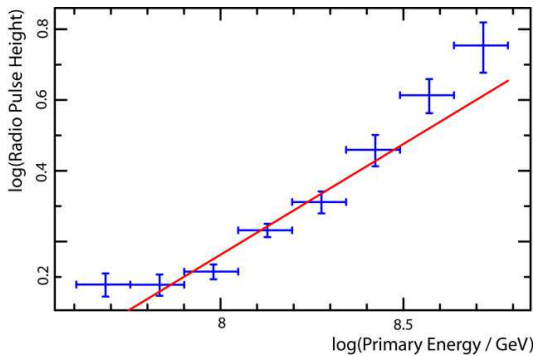}
\end{center}
\caption{Average radio pulse strength measured by LOPES in coincidence with KASCADE-Grande 
versus the estimated primary particle energy~\cite{Haungs08}.}
\label{LOPESene}
\end{figure}

Most accurate information on individual events is available from the LOPES experiment~\cite{Haungs08}. 
The main part of LOPES is an array of 30 short dipole antennas, 
which measure in coincidence with and are triggered by the well-calibrated KASCADE-Grande 
air-shower experiment. KASCADE-Grande operates in the energy range of $10^{16}-10^{18}\,$eV.  
With LOPES, some important characteristics and dependencies of the radio signal have been investigated: 
The most interesting result is the found quadratic dependence of the 
receiving radio-power on the primary energy of the cosmic particles 
(i.e.~a linear dependence on the electric field strength proving the coherence of 
the emission mechanism, fig.~\ref{LOPESene}). 
In addition, the correlation of the radio field strength with the direction of the 
geomagnetic field and the exponential behavior 
of the lateral decrease of the field strength with an scaling parameter
in the order of hundreds of meter have been explored. 
Except during strong thunderstorms
the radio signal is not strongly affected by weather conditions.  
LOPES has summarized these findings in an analytical expression for the radio pulse height 
based on the estimated shower observables:

{\small 
\begin{eqnarray}
\nonumber
&\epsilon_{\rm est}  =
(11\pm1.)
\left((1.16\pm0.025)-\cos\alpha\right) \cos\theta & \nonumber \\
\nonumber
& \exp\left(\frac{\rm -R_{SA}}{\rm (236\pm81)\,m}\right)
\left(\frac{\rm E_{p}}{\rm 10^{17}eV}\right)^{(0.95\pm0.04)} 
\left[\frac{\rm \mu V}{\rm m\,MHz}\right]&
\label{eq:horneffer-energy}
\end{eqnarray}
(With: $\alpha$ the geomagnetic angle, $\theta$ the zenith angle,
${\rm R_{SA}}$ the mean distance of the antennas to the shower axis,
and ${\rm E_{p}}$ the primary particle energy.) 
}

In particular, the quadratic dependence of the radio power on primary energy 
makes radio detection to an efficient and less costly method for measuring 
air showers. 

\begin{figure}
\begin{center}
\includegraphics*[width=7cm]{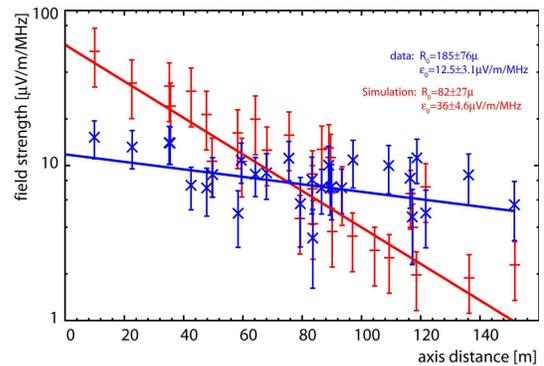}
\end{center}
\caption{Lateral distribution obtained for data
and simulation for an individual event
measured by LOPES~\cite{Haungs08}.
}
\label{LOPESlat}
\end{figure}

The high-quality data of the combined LOPES -- KASCADE-Grande measurements allow
to investigate the radio signal not only in average, but also in individual events.
In particular, the lateral distribution of the radio field strength is of importance.
Large scaling parameters would allow to measure the same field strength at
larger distances from the shower core, which could be very helpful for
large scale applications of the detection technique.
Fig.~\ref{LOPESlat} shows one individual event and compares the measured field strength
with detailed simulations performed with REAS2 on basis of the general event characteristics 
obtained by KASCADE-Grande~\cite{Nehls08}.
By investigating many events LOPES has found a good agreement in the absolute field strength 
(at $\approx 75\,$m distance from the shower axis) of simulated and measured values. 
But, regarding the lateral slope of the signal a disagreement is evident: 
The simulation of the geosynchrotron emission leads to steeper lateral slopes than the 
measurements show. 
In addition, a few percent of showers with exceptionally flat lateral distributions 
have been observed in the measurements, but they have been never found in simulations.

These results are confirmed by measurements of the CODALEMA 
experiment~\cite{Lautridou08} and 
(with much lower statistical accuracy) also by first shower detections 
at the Pierre Auger Observatory in Argentina~\cite{Coppens08}. 
CODALEMA is an experiment for observing electrical transients associated 
with cosmic rays and is located at the astronomical decametric 
antenna array at Nancay, France. 

The great success of these measurements results in an enormous gain of knowledge 
on the behavior of the radio signal in air showers. That leads to a new level of the 
studies with promising developments, but with a vast number of new questions
which necessarily have to be answered before the technique can be used for large scale 
applications:
\begin{itemize}
\item Lateral shape of the radio signal? 
All experiments have found a few percent of showers with 
exceptionally flat lateral distribution, which is not understood at all.
\item Polarization of the signal?
First hints of a dependency of the polarization on the azimuth angle of 
the incoming primary particle was found~\cite{Isar08,Lautridou08}. 
This have to explored in much more detailed as it is important for a verification of 
the favored geosynchrotron emission process. 
\item Frequency band? Is there a more efficient frequency range in the emission as the 
mainly explored $40-80\,$MHz. 
\item Geomagnetic effects? Despite the fact that there is an geomagnetic origin of the 
radio emission, the detailed dependency is still unclear.
\item Are there additional emission mechanisms like a contribution of Cherenkov emission in air?
\item Structure of the radio wavefront and the signal pulse? 
For a efficient use of the detection technique in terms of estimating primary cosmic ray
characteristics like energy, composition, or arrival direction, the wavefront and the signal pulse 
shape have to be understood in detail. 
\end{itemize}
By continuing the present measurements, but with increased quality and statistical accuracy there is 
good hope to answer these questions in near future.

Beside the understanding of the physics of the radio signal, in parallel, technical
improvements are needed for a radio stand-alone large scale application. 
First of all a working and stable self-trigger system has to be developed, 
i.e.~the antenna system itself has to recognize when a high-energy air-shower is reaching 
the observation level, but should suppress man-made RFI signals. 
Various concepts are presently under development~\cite{Revenue08,Gemmeke08}.
Also of high importance is the development of self-sustained radio stations with wireless 
communications to operate them in remote landscapes~\cite{Coppens08,Revenue08}. 
\begin{figure}
\begin{center}
\includegraphics*[width=6cm]{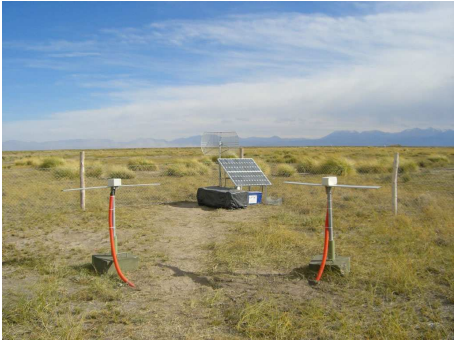}
\end{center}
\caption{At the site of the Auger Observatory in Argentina self-sustained radio stations are 
tested measuring in coincidence with the surface detectors of Auger~\cite{Revenue08}.
}
\label{rauger}
\end{figure}

\begin{figure}
\begin{center}
\includegraphics*[width=6cm]{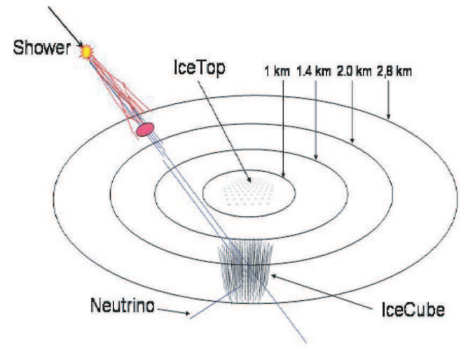}
\end{center}
\caption{Proposed extension of the IceTop array at the South Pole by radio 
antennas~\cite{Auffenberg08}.
}
\label{ricetop}
\end{figure}

Due to the success of LOPES and CODALEMA the next generation of experiments to measure 
cosmic rays by their radio emission is already in set-up or planned:
\begin{itemize}
	\item Pierre Auger Observatory: In next couple of years an engineering and testing array 
	in Argentina will be set-up (a prototype is shown in fig.~\ref{rauger}). 
	The main goal is to establish and further develop the technique and to understand the 
	radio signal in the primary energy range of $10^{17}-10^{20}\,$eV~\cite{Coppens08}. 
	\item LOFAR: The core of this radio astronomy array presently set-up in the Netherlands 
	will be able to investigate individual air showers in great detail by 
	hundreds of antennas measuring simultaneously the air-shower events~\cite{Horneffer08}.
	\item IceCube/IceTop at the South Pole: With R-IceTop a large area surface radio 
	air-shower detector is proposed, organized in rings around the IceTop array~\cite{Auffenberg08}. 
	The array will serve mainly as a veto detector for very high energy cosmic neutrinos 
	measured by the IceCube optical detectors deep in ice (fig.~\ref{ricetop}), 
	but will also provide complementary information to the air-shower array IceTop.
	\item 21CMA: The proposal is to beam the existing 21CMA radio antenna array in China to 
	to a nearby mountain in order to observe Earth skimming or through-mountain neutrino 
	trajectories resulting in a visible air shower in front of the mountain~\cite{Ardouin08}. 
\end{itemize}

\subsection{Neutrinos}

The Auger Collaboration has shown that for a certain energy range neutrino induced showers 
can be identified when studying extremely horizontal air showers.
Therefore it is of special interest to which extent a radio antenna array is sensitive to such 
horizontal air showers. As radio synchrotron signals are beamed into
the forward direction, for horizontal or earth-skimming events strong
signals can be expected and have already been seen~\cite{Saftoiu08}. 
A clear disadvantage is that most of the man-made background is also coming from the horizon.
In addition, at the frequencies used for radio detection interferences originating from 
atmospheric conditions (e.g. thunderstorms) can be propagated over
large distances through the atmosphere.  Thus a precise monitoring 
of the atmospheric conditions appears to be indispensable.

\subsection{Gamma rays}

Primary high-energy gamma rays behave in the atmosphere nearly similar to nuclei. 
The question is, if the radio signal of these more or less pure electromagnetic 
gamma induced showers is different from such 
initiated by nuclei with a considerable amount of muons?
First comparative simulations in frame of the geosynchrotron mechanism 
have shown that indeed the absolute amplitude and the lateral slope parameter of the 
radio signal is composition sensitive~\cite{huege08}. 

\section{Detection of neutrino initiated showers in solids}

Whereas in air a shower extends over many kilometers, in dense dielectric
solids the shower is much more concentrated ($\approx 10\,$cm). 
But still the number of generated $e^+ e^-$ pairs  scales with the primary energy 
of the shower initiating particle. 
In dense media the positrons drop out and the electrons scatter resulting in a charge 
imbalance of approximately 20\%~\cite{Zas06}. 
The conditions of this dense, charged pancake of relativistic particles is 
such that a coherent Cherenkov radiation is emitted in a frequency range of a few 
hundred MHz up to a few GHz. The production of a coherent RF signal out of this shower 
is named after Askaryan, who has first described the effect in 1962~\cite{Askaryan62}.
The effect could be verified in laboratory tests, where a dense particle beam was 
injected in a sand, salt, or ice target. By a variety of radio antennas the radio signal could be
detected and the strength of this signal was in the same order as expected from calculations 
of the Askaryan effect (see e.g.~ref.~\cite{Gorham05}). 

Charged particles and gamma rays interact in the atmosphere. Hence, neutrinos only 
are able to generate electromagnetic showers in solids on Earth. The large attenuation lengths of 
radio waves in dielectric materials enable to equip large detector volumes with sensors,
which is necessary because of the low interaction cross-sections of neutrinos. 
The energy threshold of a neutrino for generating a detectable radio signal is 
still an open question. 

\subsection{Antarctic ice as target}

\begin{figure}
\begin{center}
\includegraphics*[width=7cm]{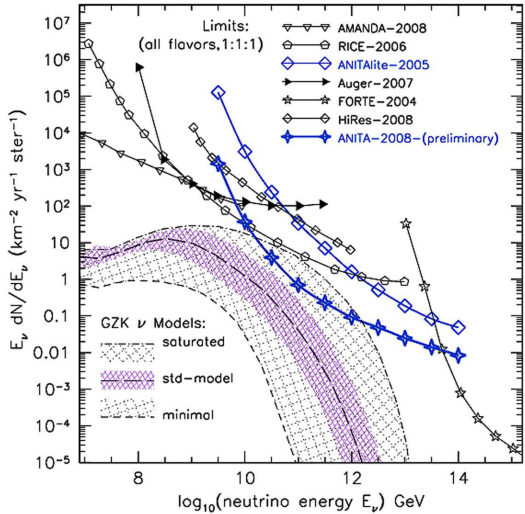}
\end{center}
\caption{Experimental upper limits on neutrino flux compared with a selection of model 
calculations for the flux of GZK neutrinos~\cite{Connolly08}.}
\label{amylimits}
\end{figure}

The path-finder of in-ice radio experiments was RICE installed on site of the optical 
neutrino experiment AMANDA at the South Pole. RICE was for 
many years the only active experiment searching for radio signals induced by 
cosmic particles. It has
taken data since 1999, and was stopped in 2005~\cite{Kravchenko06}. 
RICE measured in the frequency range of $200-500\,$MHz and could set 
limits to the diffuse neutrino flux in the primary energy range of $10^{18}-10^{21}\,$eV 
(see fig.~\ref{amylimits}). Recently the data from RICE were used 
to set an upper bound limit in the flux of highly relativistic magnetic monopoles 
($<10^{18}\,$cm$^{-2}$s$^{-1}$sr$^{-1}$)~\cite{Hogan08}.  

\begin{figure}
\begin{center}
\includegraphics*[width=7cm]{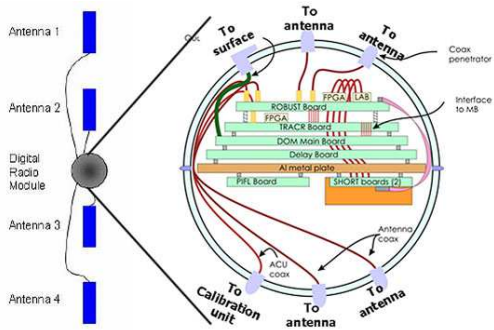}
\end{center}
\caption{Sketch of an AURA antenna cluster with 4 antennas and the Digital Radio Module (DRM) containing the electronics~\cite{Landsman08}.}
\label{aura}
\end{figure}

The concept of RICE is used for a next generation experiment of in-ice radio detection 
of neutrinos, the Askaryan Under-ice Radio Array: AURA~\cite{Landsman08}, where AURA is 
a project within the IceCube collaboration. 
A so-called AURA antenna cluster consists of 4 broad band antennas centered at 400 MHz 
and an electronic box which is adapted to the usual optical 
IceCube module in order to have a similar deployment in ice (fig.~\ref{aura}). 
Already two clusters are deployed in the ice and further three are 
foreseen to be deployed in this season. 
The present studies concentrate on the understanding the noise environment at the South Pole 
and to optimization of the hardware.

AURA will serve as prototype for a future large area radio and acoustic hybrid array centered around 
the optical IceCube. First simulations with different layouts of such an super-hybrid neutrino 
detector are performed with promising expectations on the number of detectable 
GZK-neutrinos~\cite{Tosi08}. 
Even a few events with coincide detection by two techniques can be expected, despite the very 
different threshold energies and sensitivities.
\begin{figure}
\begin{center}
\includegraphics*[width=7cm]{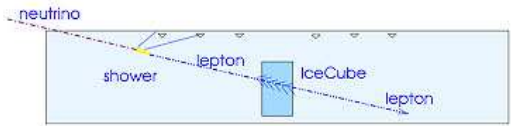}
\end{center}
\caption{Concept of IceRay, i.e.~a large antenna array in ice at shallow depths with the 
possibility to detect hybrid events with a simultaneous optical and radio detection of 
neutrinos~\cite{Kelley08}.}
\label{iceray}
\end{figure}

Another new concept to extend (the optical) IceCube into the EeV range by use of a radio array 
is IceRay~\cite{Kelley08}. The idea is to go to lower frequency ($<500$MHz), where the Askaryan 
emission happens under larger Cherenkov angles and place an antenna system at shallow depths of 
$50-200\,$m in the ice. 
Due to the large attenuation length in ice and the wide Cherenkov cone 
this would allow to increase the sensitive detector volume 
(fig.~\ref{iceray}) and to measure a substantial rate of GZK-neutrinos per year. 
To a small amount there are even events which could be seen by both, the optical IceCube 
and IceRay. Such a detection possibility would be extremely helpful for the 
proof-of-principle of the detection technique and for calibration issues.
First modules of IceRay are tested in laboratories and are ready for deployment at 
the South Pole, where the goal is to cover initially an area of 50 km$^2$ which should 
be later extended to $300-1000\,$km$^2$.

\begin{figure}
\begin{center}
\includegraphics*[width=7cm]{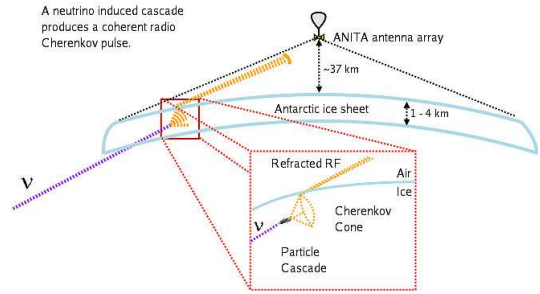}
\end{center}
\caption{Concept of the balloon borne radio experiment ANITA~\cite{Connolly08}.}
\label{anitaconcept}
\end{figure}

\begin{figure}
\begin{center}
\includegraphics*[width=3.7cm]{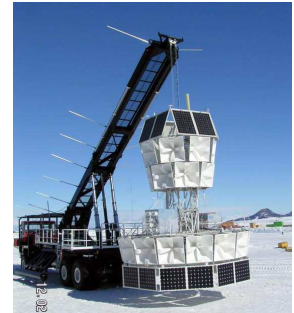}
\end{center}
\caption{The ANITA experiment ready for launch~\cite{Connolly08}.}
\label{anitaphoto}
\end{figure}

Compared to the new AURA and IceRay projects, ANITA~\cite{Barwick06} 
is a well established experiment.
The concept of the balloon borne ANITA experiment (fig.\ref{anitaconcept}) is to view from an 
elevation of $38\,$km during a long duration balloon flight a cylindrical volume of ice with 
$700\,$km in radius. 
ANITA (fig.~\ref{anitaphoto}) is equipped with 3 layers of horn antennas and register radio signals 
in the frequency range of $200-1200\,$MHz coming out of (or unfortunately reflected by) the antarctic 
ice. 
The effective threshold energy for neutrinos is at $\approx 10^{19}\,$eV, but the analysis of the 
data is hampered by the not well known transmission of the signal out of the ice.  
End of 2006 ANITA was able to perform a long flight with 18 days of good lifetime. 
For calibration a signal was used which was sent by an under-ice emitter.
A main task of the analysis concerns the polarization of the received signals as one expects a 
strongly vertical polarized signal by the Askaryan emission. 
Signals by reflections from above (e.g.~solid state relays on satellites) 
strongly favor horizontal polarization. 
No positive detection of a high-energy neutrino could be found, but
from the preliminary results of this ANITA-I flight flux limits could be estimated~\cite{Connolly08}
(fig.~\ref{amylimits}). 
These limits do constrain first (very optimistic) GZK models of the
neutrino flux, which is very promising for the future.
Data of a successful flight (foreseen in December 2008) of the slightly improved 
ANITA-II set-up will allow to start digging into the standard parameter 
space of cosmic neutrinos.

\subsection{Rock salt as target}

\begin{figure}
\begin{center}
\includegraphics*[width=6cm]{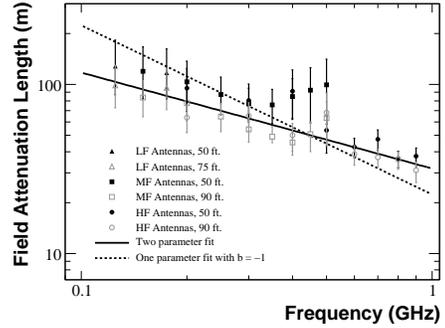}
\end{center}
\caption{Measurements of the attenuation lengths of radio signals in the Cote Blanche Salt Mine,
Louisiana, USA~\cite{Connolly08a}.}
\label{salt}
\end{figure}

Salt as target can be of advantage as it has a higher density compared to ice 
($2.2\,$g/cm$^2$ versus $0.92\,$g/cm$^2$) and the accessibility of salt domes 
appears to be less difficult than of the Antarctica. 
In addition, the soil or water above the salt domes provide a good RF insulation.
Disadvantages are probably the high drilling costs for installing radio sensors in salt, 
when a dense array is needed. This would the case if the attenuation lengths is too short. 
For an effective use of salt as target one need attenuation lengths at least in the order of 
$200\,$m to $500\,$m in the interval of $100\,$MHz to $1\,$GHz.  
Unfortunately rock salt domes seems to be very inhomogeneous and there is yet no 
clear conclusion about the attenuation of radio waves in salt. 
Figure~\ref{salt} shows results from recent measurements~\cite{Connolly08a} done in the frame 
of the SalSA project with an attenuation lengths of $51 \pm 3\,$m at $440\,$MHz 
(solid line in fig.~\ref{salt}). 
It may be stated that currently the use of salt as target appears to be hardly 
competing with the sensitivity of radio neutrino detectors in ice. 
However, it is  certainly worthwhile to explore the possibilities of any medium
which can complement the antarctic ice in the sense of a northern accessibility 
to ultra-high energy neutrinos. 

\subsection{Permafrost as target}

Recently, a discussion was initiated to use permafrost as target, 
where neutrino induced showers can be detected by acoustic and radio sensors~\cite{Nahnhauer08}. 
Permafrost covers $\approx$20\% of land on Earth.
Its thickness extends to more than $500\,$m and it
is located in the northern hemisphere in Siberia.
To test the characteristics of the medium it is proposed to install TAIGA, a Test Askaryan 
Installation for Geology and Astrophysics.

\section{Radio detection of particles impinging the moon}

As already proposed by Askaryan, the moon provides a convenient, radio-quiet and huge target 
in the sky. 
Neutrinos from the cosmos - and due to the absence of an atmosphere additionally photons 
and high-energy particles - do collide with the moon rock in the regolith and a coherent 
radio pulse is emitted which escape through the surface and may be detected on Earth. 
However, the threshold of the primary energy for a detection is even higher than for radio detection in ice. 
First measurements 
(e.g. with the Goldstone dish: GLUE~\cite{Gorham04}) result in flux limits for energies 
in the range of $10^{21}-10^{24}\,$eV. Probably, we do not expect to have any neutrinos at such high 
energies, but there is hope that with improved techniques the sensitivity will be extended to lower energy. 

The strategic way is to optimize already existing astronomical radio telescopes 
or radio detection experiments currently in preparation, rather than setting up completely new dedicated arrays. The possibility to add a further use to existing installations has certainly triggered the world wide interest in measurements of radio signals from the moon.

\begin{figure}
\begin{center}
\includegraphics*[width=8cm]{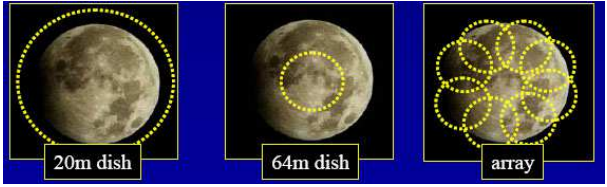}
\end{center}
\caption{With an array of radio antennas the entire limb of the moon can be 
observed~\cite{Ekers08}.}
\label{moonarray}
\end{figure}

During the conference there have been interesting discussions on substantial details related to the 
optimum strategy of moon observations in order to detect first ultra-high energy neutrinos:
\begin{itemize}
	\item Frequency range: The Askaryan emission in the GHz range has a very defined and sharp 
	Cherenkov angle around $60^\circ$. Therefore, the sensitivity increases at lower energies 
	going to higher frequency (for constant pulse detection threshold of the telescope). 
	On the other hand, a wide spread Cherenkov emission, as expected at MHz frequencies, 
	allows to observe a much larger volume of the target, 
	i.e.~at lower frequencies the volume of the detector increases, 
	and therefore the flux sensitivity improves~\cite{Scholten08}.
	\item Advantage of antenna arrays: Due to the Cherenkov like emission of 
	the radio signal the event rates should be highest at the limb of the moon. The noise contribution 
	is the same from all the moon. Therefore, the optimum would be to observe the entire limb of 
	the moon, but not the center~\cite{Ekers08}. This is only possible by using an array of 
	antennas instead of a single dish (see fig.~\ref{moonarray}).
	\item Imaging array: An array of simple and digital antennas (as foreseen for LOFAR~\cite{Falcke08}
	 and SKA) would have the advantage of performing multiple beam observations at the 
	 same time. By that there will be enough beams to cover all the moon limb~\cite{Ekers08}.
	\item Dispersion of the signal: To improve the separation capability of signal to RF noise 
	the dispersion of the detected signal can be used. Terrestrial noise should 
	have no dispersion, terrestrial signals bounced on satellites should have double dispersion, 
	but signals from the moon should have a characteristic dispersion due to the 
	known distance~\cite{James08}.
	\item Regolith characteristics: An important (still) unknown factor in calculating the 
	sensitivities is related to the thickness of the lunar regolith and irregularities of its 
	density and surface.  
	Useful indications will be obtained by the LORD mission, a Lunar Orbital Radio Detector. 
	The goals of LORD~\cite{Gusev08} are generally the study of the moon and its vicinity 
	(e.g.~EM environment, plasma, lunar seismic activity) and in particular the detection 
	of particles of energies above $10^{20}\,$eV. 
\end{itemize}
\begin{figure}
\begin{center}
\includegraphics*[width=6cm]{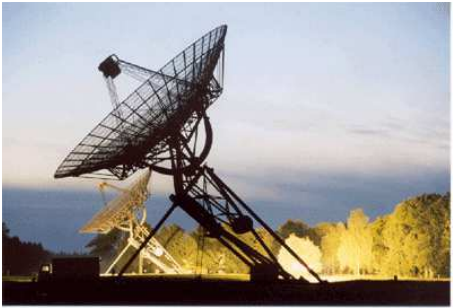}
\end{center}
\caption{The Westerbork Synthesis Radio Telescope (WSRT) used for radio Moon observations.}
\label{westerbork}
\end{figure}

\begin{figure}
\begin{center}
\includegraphics*[width=7cm]{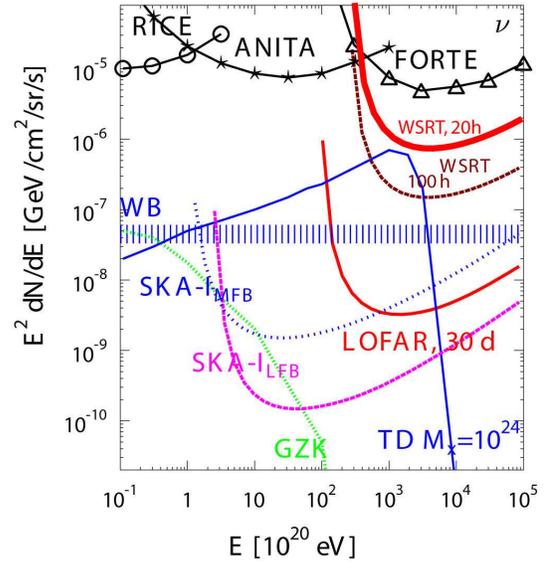}
\end{center}
\caption{Upper limits on neutrino flux as a function of neutrino energy~\cite{Scholten08}. 
Displayed are experimental limits, expectations on limits for future experiments, as well as
the calculated Waxmann-Bahcall limit and expectations on the neutrino flux from 
topological defects and of GZK-neutrinos.}
\label{skan}
\end{figure}

Most experimental progress in the field is by the NuMoon~\cite{Scholten08} experiment.
First results are available from an analysis of 20 hours of observation 
using the Westerbork Synthesis Radio Telescope (WSRT, fig.~\ref{westerbork}) 
searching for short radio pulses from the moon.
WSRT, an telescope array, consists of 14 parabolic telescopes of $25\,$m dishes 
on a $2.7\,$km east-west line. For the moon observation the frequency range 
115-180$\,$MHz were used and full polarization data were recorded in two beams 
of 4 narrow bands each. The two beams are aimed at different 
sides of the Moon, each covering about one third of the lunar surface. 
A real lunar Cherenkov pulse should be visible in only one of the two beams. 
The obtained limit on the neutrino flux is shown in fig.~\ref{skan} together with the 
expectation for 100 hours of observation with WSRT, what already will rule out some 
of the models for neutrino fluxes, in particular those based on topological 
defects~\cite{Scholten08}.

A next step of moon observations by the NuMoon collaboration will be the 
use of LOFAR~\cite{Falcke08}, 
the Low Frequency Array, presently under construction in the Netherlands. 
Hundreds of low frequency omni-directional radio antennas will be located 
inside a very dense core. By that, multiple beams can be formed
to cover the surface of the moon, resulting in a sensitivity of about a factor 20 higher
than the WSRT (see fig.~\ref{skan}).

A few hours of moon observations have also been carried out with the 
Australia Telescope Compact Array (ATCA)~\cite{James08}, but with a much lower sensitivity 
compared to NuMoon. 
The LUNASKA project  at ATCA experiences currently an upgrade towards measurements 
with a bandwidth of $2\,$GHz~\cite{James08}. 

In future, the best sensitivity will be achieved with the Square Kilometer
Array (SKA)~\cite{Ekers08,Scholten08} planned to be completed in 2020. 
In fig.~\ref{skan} the expected sensitivity of SKA is shown for observations 
in the low frequency band (70-200$\,$MHz) and the middle frequency band (200-300$\,$MHz).

\begin{figure}
\begin{center}
\includegraphics*[width=7cm]{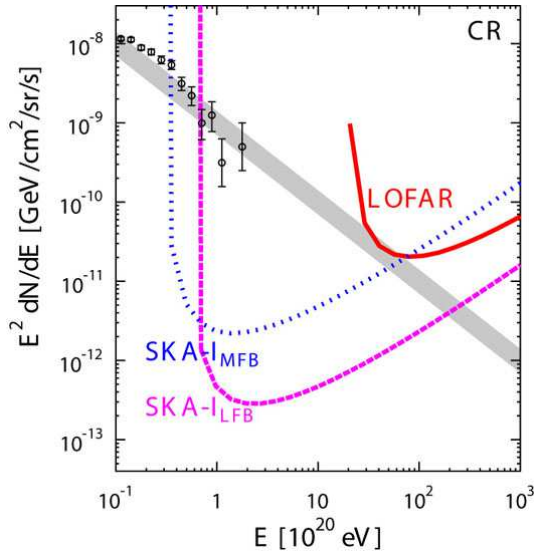}
\end{center}
\caption{Expected sensitivities of lunar radio observations to charged cosmic rays with 
the LOFAR and SKA (low and middle frequency bands) antenna arrays~\cite{Ekers08,Falcke08}.}
\label{skah}
\end{figure}

As already indicated, moon observations of radio signals generated by the Askaryan-effect 
are also sensitive to charged cosmic rays and UHE photons. 
An optimized SKA array~\cite{Ekers08,Scholten08} expects to have a sensitivity to scratch the 
guaranteed charged cosmic ray flux (see fig.~\ref{skah}). 
Of course that would be not only a great success, but also a clear 
proof-of-principle of the lunar radio observation technique. 

\section{Perspectives}

Radio experiments have made enormous progress within the last five years. 
The initial experiments for both air shower observations and searching for 
neutrino signals in dense media were technically successful and are ready for 
an application in next generation experiments.

Radio detection of air showers with the advantage of coincidence measurements of 
the same showers with other techniques has reached in the meanwhile a good 
understanding of the received signal and its dependence on the the shower parameters. 
This is true at least for primary energies $<10^{18}\,$eV. 
The next generation of experiments (Radio at the Pierre Auger Observatory or LOFAR) will 
concentrate on precision measurements of the signal to verify the emission mechanism and on the 
engineering of radio stand-alone, large-scale experiments. In particular, the ability to 
self-trigger on the radio signal and to develop robust self-sustained radio detector 
stations. However, in summary, already the present results provide strong arguments in 
favor of the use of the radio technique to study the origin of high-energy cosmic rays.

For neutrino detection by radio wave generation in dense media 
also a hybrid device based on different 
detection technologies (e.g.~radio, acoustic, and optical)
would be the most powerful realization for a proof-of-principle of the technique. 
But due to the different energy thresholds and sensitivities of optical, radio,
and acoustic detection such task is much more difficult than in case of air showers. 
Nevertheless, currently several groups are working toward the realization of such an 
observatory at the South Pole.  Due to the inability to propagate radio 
waves through water or light through rock salt, the South Pole is most likely the 
only place on Earth suitable  for such an effort. 

The very promising improvements of the lunar radio observation technique will result in 
complementary information on both, neutrino physics and on the development of the 
radio detection technique of cosmic particles. 

In addition, further detector 
materials and techniques will be explored in near future. 
For all applications of radio detection of particles from the cosmos still 
dedicated R\&D projects are going to get performed. 
The next ARENA conference with new results is eagerly awaited. 

\ack{
The author would like to thank the organizers of ARENA to provide the frame for very fruitful 
discussions during the conference. 
Heinigerd Rebel is acknowledged for proofreading the manuscript.
I am indebted to all participants of the conference and to all others currently 
active in the field. I apologize to all colleagues whose presentations or work could not 
be considered in the actual summary. 
}

\end{document}